\documentclass[prl,twocolumn,showpacs]{revtex4}
\usepackage{graphicx}
\def\Journal#1#2#3#4{{#1} {\bf #2}, #3 (#4)}
\def\PR{Phys. Rev.}
\def\PRL{Phys. Rev. Lett.}

\def\PRB{Phys. Rev. B}

\def\PhysicaA{Physica A}
\newcommand{\n}{\nonumber}
\newcommand{\bn}{\begin{eqnarray}}
\newcommand{\en}{\end{eqnarray}}
\newcommand{\h}{\hspace}
\begin{document}
\title {Controlled Flow of Spin-Entangled Electrons via
Adiabatic Quantum Pumping}
 \author{Kunal K. Das, Sungjun Kim, and Ari Mizel}
 \affiliation{Department of Physics, The Pennsylvania State
University, University Park, Pennsylvania 16802}

\date{\today }
\begin{abstract}

We propose a method to dynamically generate and control the flow
of spin-entangled electrons, each belonging to a spin-singlet, by
means of adiabatic quantum pumping.  The pumping cycle functions
by periodic time variation of localized two-body interactions. We
develop a generalized approach to adiabatic quantum pumping as
traditional methods based on scattering matrix in one dimension
cannot be applied here. We specifically compute the flow of
spin-entangled electrons within a Hubbard-like model of quantum
dots, and discuss possible implementations and identify parameters
that can be used to control the singlet flow.

\end{abstract}
 \maketitle

Entanglement is one of the most intriguing features that
distinguish the quantum world from the classical. In recent years
\cite{Nielsen-Chuang} applications potential in quantum
information and computation has injected renewed vigor into the
study of entangled states. Since Bohm's \cite{Bohm} reformulation
of the Einstein-Podolsky-Rosen (EPR) paradox \cite{EPR} in terms
of pairing of spins, spin singlets have become the canonical
example of an elementary entangled state.  From an experimental
standpoint, manipulating singlet pairs of electron spins, in
particular, is promising because of the vast expertise already
available in solid-state electronics \cite{Awschalom}.

Several proposals have emerged
\cite{Oliver,Saraga-Loss,Hu-Sarma,cooper-pairs} that aim to
generate a controlled flow of electron spin singlets using Coulomb
blockade and tunnelling at pinched quantum dots that have weak
coupling to leads. In these proposals, the natural tunnelling of
unwanted single electrons can be difficult to suppress and/or
singlet delivery time is constrained by tunnelling rates that are
slow due to the necessity for pinched dots. Since manipulations in
a quantum information device need to occur before decoherence sets
in, fast delivery is essential, and any background of unpaired
electrons reduces the purity of entanglement. It is therefore
highly desirable to have a fast dynamic mechanism to generate a
selective flow only of electrons belonging to singlets. In this
paper we show that this can be accomplished through a generalized
adiabatic quantum pumping that is induced by localized
interactions and we discuss possible ways of experimental
realization. Pinched dot tunnelling times and singlet formation
times do not constrain delivery rates in our proposal. In
addition, the absence of bias and the adiabatic nature of the
pumping can reduce the production of heat in the apparatus.

\begin{figure}[b]\vspace{-1.6cm}
\hspace{-1cm}\includegraphics*[width=\columnwidth,angle=-90]{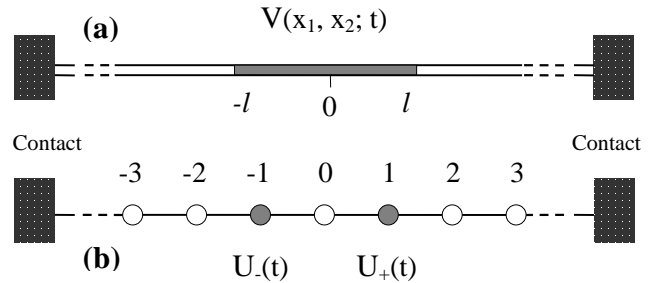}\vspace{-3.8cm}
\caption{(a) Schematic figure showing the \emph{two}-body
interaction $V(x_1,x_2;t)$ acting in a finite interval $(-l,l)$ in
a one dimensional system, (b) Implementation in a tight-binding
model where the two-body interaction is present only at lattice
sites $m=\pm 1$; the interaction strengths $U_{\pm}(t)$ at the two
sites are the time-dependent pumping parameters. } \label{Fig1}
\end{figure}

The notion of quantum pumping has its roots in a speculative paper
by Thouless \cite{Thouless} in 1983, but advances in nanoscale
transport have led to a renewed and growing interest in the
phenomenon in recent years both theoretically and experimentally
\cite{Brouwer,turnstile, nano-transport, Chamon, Watson,
Floquet-Buttiker,Floquet-Kim}. Quantum pumping is a coherent
process that creates a direct current in the \emph{absence} of any
bias through a nanoscale device, by changing its scattering
properties periodically through independent adiabatic variation of
two or more physical parameters. Adiabatic quantum pumping of
charge \cite{Brouwer}, spin \cite{Chamon,Watson} and thermal
\cite{Floquet-Buttiker} currents have been considered. However
previous studies have generally relied on a theoretical
description based on transmission and reflection coefficients
which cannot be used to describe the pumping of singlets due to
localized interactions; therefore in this paper we also formulate
a more generalized approach to the theory of quantum pumping.

\emph{Singlet Current}:  We consider a single available channel in
a quasi one dimensional mesoscopic conductor connected to
macroscopic contacts (Fig.~\ref{Fig1}(a)). In the presence of two
body interactions the charge current can be defined in terms of
the two particle reduced density matrix, $\rho_2$. If the
interaction does \emph{not} affect spins, the reduced density
matrix separates into four independent spin subspaces, one singlet
with a symmetric spatial part and three triplets  with
antisymmetric spatial parts; the singlet current is therefore
equivalent to the charge current associated with the symmetric
spatial part ($\rho_2^{S}$) of the two particle reduced density
matrix
\bn  J_S(x_1,t)=\frac{e\hbar}{m}\int dx_2
\Im\left\{\partial_{x_1}\rho_2^{S}(x_1,x_2;x_1',x_2;t)\right\}_{x_1=x_1'}.\en
We consider a situation in which two-body interactions are
localized, meaning that they are non-vanishing only when the
particles are in a certain finite interval, as shown in
Fig.~\ref{Fig1}(a). Such interactions lend themselves to a
scattering description and the current can be approximately
evaluated by expanding the density matrix in terms of two particle
scattering states
\begin{widetext}
\bn\label{current} \h{-1mm}J_{S}(x_1,t)\simeq\frac{e\hbar}{2m}\int
\h{-1mm}dE\ F(E)
\h{-1mm}\int\frac{dk_1}{2\pi}\h{-1mm}\int\frac{dk_2}{2\pi} \
\delta
\left({\textstyle\frac{\hbar^2k_1^2}{2m}+\frac{\hbar^2k_2^2}{2m}}-E\right)
\h{-1mm}\int dx_2\Im\left\{\partial_{x_1}
\Psi_{k_1,k_2}(x_1,x_2,t)\Psi_{k_1,k_2}^*(x_1',x_2,t)\right\}_{x_1=x_1'}.
\en
\end{widetext}
Here $E$ denotes the energy required to remove a pair of particles
from the many-body ground state, and $F(E)$ is the distribution of
this pair energy. The effect of interaction on the current is
determined completely by the two-particle \emph{singlet}
scattering states, $\Psi_{k_1,k_2}$ arising from \emph{free}
singlet states
$\Phi_{k_1,k_2}(x_1,x_2)=\frac{1}{\sqrt{2}}[\phi_{k_1}(x_1)\phi_{k_2}(x_2)+
\phi_{k_1}(x_2)\phi_{k_2}(x_1)]$ where $\phi_{k}(x)$ denotes a
single particle plane wave state with momentum $\hbar k$.

\emph{Pumped Current through Adiabatic Perturbation }:
The two particle scattering states, and therefore the current, are
determined by the interaction $V(\bar{x},t)$ between a pair of
particles; we take it to be time-dependent and to occur only in a
\emph{finite} region $|x_i|<l$. Most importantly the interaction
$V(\bar{x},t)$ is chosen to be localized so that it \emph{only
affects singlets thereby naturally eliminating the flow of
triplets} in the absence of a bias. When the characteristic period
$\omega$ of the time variation of the potential is slow compared
to the time $\delta t$ the particles dwell in the scattering
region \cite{Buttiker-Landauer}, $\omega\times\delta t\ll 1$, we
can apply adiabatic perturbation theory to express the scattering
states of the time-dependent Hamiltonian in terms of the
instantaneous states up to linear order
\bn\label{wf-lin-ord}\Psi_{\bar k}(\bar{x},t)\simeq\Psi^t_{\bar
k}(\bar{x})-i\hbar\int d\bar{x}'
G^t(\bar{x},\bar{x}';E)\frac{\partial}{\partial t}\Psi^t_{\bar
k}(\bar{x}').\en
We use the notation $\bar{x}\equiv \{x_1,x_2\}$ and $\bar{k}\equiv
\{k_1,k_2\}$, so that $G^t(\bar{x},\bar{x}';E)$ is the
\emph{two}-particle \emph{instantaneous} retarded Green's function
for the \emph{full} Hamiltonian. The instantaneous state
$\Psi^t_{\bar k}(\bar{x})$ is a solution of the
\emph{time-independent} Lippmann-Schwinger equation for the
potential $V(\bar{x}',t)$ at the specific time $t$,
\bn\label{LS} \Psi^t_{\bar k}(\bar{x})&=&\Phi_{\bar
k}(\bar{x})+\int d\bar{x}'
G_0(\bar{x},\bar{x}';E)V(\bar{x}',t)\Psi_{\bar k}(\bar{x}').\en
where $G_0$ is the \emph{ free two-particle retarded} Green's
function. Taking the time derivatives of the defining equations
for $\Psi^t(\vec{x}')$ and $G^t(\bar{x},\bar{x}';E)$ enables us to
express the second term in Eq.(\ref{wf-lin-ord}), that is linear
in $\partial_t$,  as
\bn \Delta\Psi_{\bar k}(\bar{x},t)=-i\hbar\int d\bar{x}'\int
d\bar{x}'' G^t(\bar{x},\bar{x}';E)\h{1.6cm}
\\
\times G^t(\bar{x}',\bar{x}'';E) \dot{V}(\vec{x}'',t)\Psi^t_{\bar
k}(\vec{x}''). \n\en

If there is no bias or time-dependence, the laws of thermodynamics
demand that there should be no current; we explicitly confirm that
our expression for the current satisfies this essential physical
requirement. The net current in the absence of time-dependence is
evaluated by using the zeroeth order term from
Eq.~(\ref{wf-lin-ord}) for the scattering state in
Eq.~(\ref{current}):
$\Psi_{k_1,k_2}(x_1,x_2,t)\equiv\Psi^t_{k_1,k_2}(x_1,x_2)$. The
resulting expression can be simplified by relating the imaginary
part of the retarded Green's function to the free singlet states:
$\Im\{G_0(\bar{x},\bar{x}';E)\}=-\pi\int dk_1\int dk_2
\delta\left({\textstyle\frac{\hbar^2k_1^2}{2m}+\frac{\hbar^2k_2^2}{2m}}-E\right)
\Phi_{\bar{k}}(\bar{x}) \Phi^{*}_{\bar{k}}(\bar{x}')$.
Then repeated use of the Lippmann-Schwinger equation and
properties of the Green's functions shows that the net current
corresponding to the zeroeth order $\Psi^t(\bar{x})$ vanishes.

After confirming that our expression cannot produce spontaneous
current, we evaluate the singlet current induced by the adiabatic
time evolution. To linear order in the time dependence this
involves the evaluation of
\bn \int dx_2\Im\left\{\partial_{x_1} \Psi^t_{\bar k}(x_1,x_2)
\Delta\Psi^{*}_{\bar k}(x'_1,x_2,t) \right\}_{x_1=x_1'} \en
within the expression for the current in Eq.~(\ref{current}).  A
calculation employing standard Green's function identities,
similar to that for the zeroeth order, leads to an expression for
the net amount of singlet entangled electron pairs pumped in a
complete cycle of period $\tau$, yield the main result of this
paper:
\begin{widetext}
\bn Q_S(\tau) &=&\frac{e\hbar^2}{2\pi m}\int_0^\tau dt\int dE \
F(E)\frac{\partial}{\partial E}\left[\int
d\bar{x}'\dot{V}(\bar{x}',t) \int dx_2
\Im\{G^{t*}(\bar{x},\bar{x}';E)
\partial_{x_1}G^t(\bar{x},\bar{x}';E)\}\right]. \label{pumped-current}\en\end{widetext}

\emph{Singlet Pumping in a Turnstile Model}:
We illustrate our results with a tight-binding model,
(Fig.~\ref{Fig1}(b)) with two Hubbard impurities located at sites
$-m, m$
\bn V(\bar{n},t)=U_-(t)\delta_{n_1,-m}\delta_{n_2,-m}+
U_+(t)\delta_{n_1,m}\delta_{n_2,m}.\en
An electron can interact with another only at those two sites
therefore, due to the Pauli principle, only singlets are affected.
The strength of the interactions, $U_\pm(t) $ are the two
time-dependent pumping parameters. This concept is similar to a
`turnstile model' \cite{turnstile} but differs significantly in
that, instead of time-varying external potentials, the two-body
interaction among electrons is varied in time. We separately
derived a discrete version of Eq.~(\ref{pumped-current}); the end
result amounts to replacing the coordinate arguments with site
indices $x\rightarrow n$, integrals by sums and derivatives with a
finite difference form. A lengthy calculation leads to an
expression for the singlets pumped in a complete cycle in terms of
the \emph{free two-particle lattice} Green's function,
specifically two of its matrix elements
$G_0(0)=G_0(\bar{m},\bar{m};E)$ and
$G_0(2\bar{m})=G_0(\bar{m},-\bar{m};E)$
\begin{widetext} \bn  Q_S(\tau)=\frac{-e}{2\pi}\int_0^\tau\h{-2mm}dt
\h{-1mm}\int \h{-1mm}dE F(E)\frac{\partial}{\partial E}\sum_{\pm}
\dot{U}_{\pm}(t) \frac{|T_\pm(t)|^2\left[
\Im\{G_0(0)\}(1+|T_\mp(t)G_0(2\bar{m})|^2)\pm
2\Im\{T_\mp(t)G_0(2\bar{m})G_{0}^{\pm}(2\bar{m})\}
\right]}{U_\pm(t)^{2}|1-T_\mp(t)T_\pm(t)G_0(2\bar{m})G_0(2\bar{m})|^2}.\en
\end{widetext}
Here $\bar{m}\equiv\{m,m\}$, and
$T_\pm(t)=1/(U_\pm^{-1}(t)+G_0(0))$ is the T-matrix for a single
Hubbard impurity, and $G_{0}^{\pm}\equiv G_0(2\bar{m})$ for
$\lambda=\pm i|\lambda|$, when expressed in the form
$G_0(2\bar{m})=\int_{0}^{\pi}\frac{dk}{2\pi}\frac{e^{2m(ik-\lambda)}}{\sinh(\lambda)}$
with $\cosh(\lambda)=E/2-\cos(k)$.

\begin{figure}[b]
\vspace{-1.5cm}\hspace{-5mm}\includegraphics*[width=\columnwidth,angle=0]
{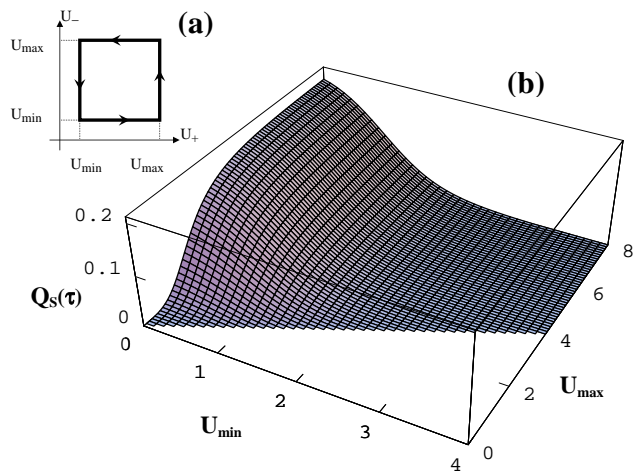} \vspace{-4.5cm} \caption{(a) Pumping
cycle in the space of parameters $U_\pm$. (b) Net singlets pumped
in a single cycle as a function of the location and size of the
square pumping cycle as $U_{min}$ and $U_{max}$ are varied.}
\label{Fig2}
\end{figure}

 Exact analytical forms exist for the lattice Green's functions, $G_0$,
in terms of elliptic integrals \cite{Economou}. For the purpose of
numerical estimates, we assume a square-profile time dependence
\cite{nano-transport} in the plane of the parameters $U_\pm(t)$,
shown in Fig.~\ref{Fig2}(a), where the two parameters change
alternately between minimum value $U_{min}$ and a maximum value
$U_{max}$. The pair distribution function is taken to be a Fermi
function, $F(E)\simeq 1/[e^{\beta(E-{\cal E})}+1]$. At low
temperatures, an integration by parts with respect to energy
yields an expression for the pumped singlets in terms of the
maximum energy, ${\cal E}$ available for a pair.

\begin{figure}[t]
\includegraphics*[width=\columnwidth,angle=0]{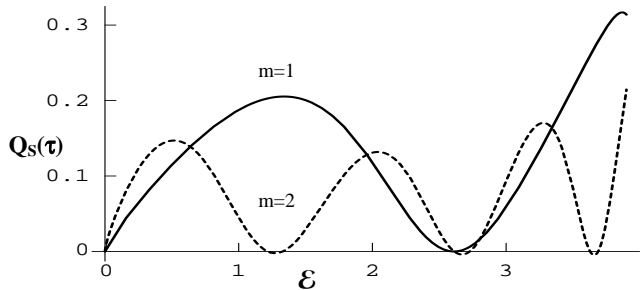}
\vspace{-7.8cm} \caption{Net singlets pumped per cycle as a
function of the maximum available energy $\cal E$ for pairs.  The
two curves correspond to different separations of the Hubbard
impurities, solid line $\rightarrow \pm m=\pm 1$ and dashed line
$\rightarrow \pm m=\pm 2$. } \label{Fig3} \end{figure}

The onsite energy of each tight-binding site is taken to be zero
and all energies are expressed in units of the nearest neighbor
coupling strength. In Fig.~\ref{Fig2}(b) we plot the net singlets
pumped in a single cycle as a function of the size and location of
the square footprint of the time-cycle in the space of the
parameters $U_\pm$; the flow depends on the enclosed region.
Figure~\ref{Fig3} shows the dependence of the singlet current on
the parameter ${\cal E}$ that measures the available energy for
pairs of electrons determined by the chemical potential in the
contacts. The two curves in the figure correspond to different
locations of the Hubbard impurities, at lattice sites $m=\pm 1$
and $m=\pm 2$, illustrating the significant effect the spatial
separation of two impurities have on the pumping rate. The
direction of flow can also reverse for certain values of the
various parameters; reversing the time-cycle is not the only way
to reverse the direction of the current \cite{Brouwer}. Quantum
pumping has the intrinsic property that the magnitude of the
pumped quantity, in this case singlets, is continuous in nature so
that the delivery rate per cycle can be continuously adjusted.
Thus there are several ways to precisely control the magnitude and
direction of the flow of singlets dynamically.

\emph{Discussion and Outlook}: The turnstile model could be
implemented by taking the interaction sites to be quantum dots
coupled to leads and varying the electron-electron interaction by
decreasing the size of the dot periodically. Concurrent variation
in the dot-lead coupling could in principle be compensated by
counter-varying gate voltages. Another way to vary the interaction
strength is to change the dielectric constant locally. This could
be accomplished by a local shift of the electron wavefunction
among layers of different materials in a heterostructure; this
technique was recently used to control the electron $g$ factor
\cite{Jiang}. A shift between silicon and germanium layers or
gallium arsenide (GaAs) and aluminium arsenide (AlAs) layers can
in principle produce up to $25\%$ change in the dielectric
constant, sufficient to see the effects described here.

The adiabatic condition requires the period, $\tau\gg\delta t$,
the dwell time given by $\delta t=d/v$, with $d$ being the size of
the scattering region and $v$ the carrier velocity
\cite{Buttiker-Landauer}. Since our singlet pump requires no
potential barriers that reduce kinetic energy, $v$ may be taken to
be the Fermi velocity typically $>10^5$ m/s, so for a scattering
region $d\sim 10$ nanometers the adiabatic condition would allow
thousands of cycles per nanosecond. Our numerical estimates then
give a pumping rate at the order of a thousand singlets per
nanosecond. In typical Coulomb-blockade based schemes the most
optimistic estimates yield a delivery rate at the order of one
nanosecond per singlet \cite{Saraga-Loss,Hu-Sarma}. Thus our
approach has the potential to be much faster.

Our result, Eq.~(\ref{pumped-current}), has the merit that it can
also be applied to quantum pumping in systems that allow an
independent particle description, as considered in previous
studies. All the elements in Eq.~(\ref{pumped-current}) then
reduce to single particle functions: $F(E)\rightarrow f(E)$ is the
Fermi distribution function, the Green's function is a
\emph{single} particle one with an asymptotic form
${\rightarrow}-i[m/(\hbar^2 k)] e^{ikx}\psi_k^{t*}(x')$  in terms
of  the scattering state $\psi_k^t(x')$ and wavevector
$k=[2mE/\hbar^2]^{-1/2}$. This yields the pumped \emph{charge}
\bn Q(\tau) =\frac{em}{2\pi\hbar^2}\int_0^\tau\h{-2mm}dt\int dE \
f(E)\frac{\partial}{\partial E}\left\{\frac{1}{
k}\langle\psi^t|\dot{V}|\psi^t\rangle\right\},\en
which agrees with expressions derived in earlier works
\cite{nano-transport, Brouwer}. But unlike all previous
treatments, we never use 1D scattering matrix elements as they do
not have useful generalizations when particles interact with each
other in a region rather than scatter off an external potential.

To summarize, we have proposed a method based on two-body
adiabatic quantum pumping for generating a dynamically controlled
flow of spin-entangled electrons.  The process is inherently
coherent, potentially much faster than most current proposals, has
reduced noise because of the lack of bias and the natural
elimination of single electron flow, and allows for continuous
adjustment of the flow through numerous physical parameters. All
of these features can be developed and incorporated into a
comprehensive scheme to generate a controlled flow of entangled
electrons. Our goal here has been to present the basic idea,
develop a theoretical framework for its description and discuss a
possible physical model for implementation.  We have in the
process generalized the treatment of quantum pumping to
incorporate interactions that cannot be treated in a scattering
matrix approach. We anticipate that future studies can build upon
the considerations in this work.

This work was supported by the Packard Foundation and NSF NIRT
program grant DMR-0103068.

\emph{Note added}: When we were finishing this paper we become
aware of an excellent recent proposal by Beenakker \emph{ et al.}
\cite{Beenakker} to excite entangled electron-hole pairs using
one-body potential in a scattering matrix approach. Our proposal
is more difficult to implement, but it could be naturally
incorporated into an electron-only spintronic device, it produces
a current of singlets unadulterated by single electrons, and it
will enjoy a longer decoherence time since hole decoherence times
tend to be short.


\begin{thebibliography}{4}

\bibitem{Nielsen-Chuang} M. A. Nielsen and I. L. Chuang,
\emph{Quantum Computation and Quantum Information}, (Cambridge,
New York, 2000).

\bibitem{Bohm} D. Bohm, \emph{Quantum Theory}, (Prentice-Hall, New
Jersey, 1951).

\bibitem{EPR} A. Einstein, B. Podolsky, and N. Rosen, \Journal{\PR}{47}{777}{1935}.

\bibitem{Awschalom}
\emph{Semiconductor Spintronics and Quantum Computation}, edited
by D.D. Awschalom, D. Loss, and N. Samarth (Springer, 2002).

\bibitem{Oliver}  W. D. Oliver, F. Yamaguchi, and Y. Yamamoto,
\Journal{\PRL}{88}{037901}{2002}.

\bibitem{Saraga-Loss}  D. S. Saraga and D. Loss,
\Journal{\PRL}{90}{166803}{2003}.

\bibitem{Hu-Sarma}  X. Hu and S. Das Sarma,
\Journal{\PRB}{69}{115312}{2004}.

\bibitem{cooper-pairs}  P. Recher, E. V. Sukhorukov, D. Loss
\Journal{\PRB}{63}{165314}{2001}.

\bibitem{Thouless}  D. J. Thouless,
\Journal{\PRB}{27}{6083}{1983}.

\bibitem{Brouwer} P. W. Brouwer, \Journal{\PRB}{58}{R10135}{1998}.

\bibitem{turnstile}  Y. Levinson, O. Entin-Wohlman, and P. W\"{o}lfle,
\Journal{\PhysicaA}{302}{335}{2001}.

\bibitem{nano-transport} O. Entin-Wohlman, A. Aharony, and Y. Levinson, \Journal{\PRB}
{65}{195411}{2002}.

\bibitem{Chamon}  P. Sharma and C. Chamon,
\Journal{\PRL}{87}{096401}{2001}.

\bibitem{Watson} S. K. Watson, R. M. Potok, C. M. Marcus, and V.
Umansky, \Journal{\PRL}{91}{258301}{2003}.

\bibitem{Floquet-Buttiker} M. Moskalets and M. B\"{u}ttiker,
\Journal{\PRB}{66}{205320}{2002}.

\bibitem{Floquet-Kim} S. W. Kim, \Journal{\PRB}{66}{235304}{2002}.

\bibitem{Buttiker-Landauer}  M. B\"{u}ttiker and R. Landauer,
\Journal{\PRL}{49}{1739}{1982}.

\bibitem{Economou} E. N. Economou, {\it Green's Functions in Quantum
Physics}, (Springer-Verlag, Berlin 1979).

\bibitem{Jiang} H. W. Jiang and E. Yablonovitch, \Journal{\PRB}{64}{041307}{2001}.

\bibitem{Beenakker}  C. W. J. Beenakker, M. Titov, and B.
Trauzettel, \Journal{\PRL}{94}{186804}{2005}.

\end{thebibliography}
\end{document}